\documentclass[twocolumn,nofootinbib,amsmath,amssymb,aps,prd,balancelastpage,superscriptaddress]{revtex4-1}

\usepackage{color}
\usepackage{xcolor}
\usepackage[active]{srcltx}
\usepackage{amsmath,amsfonts,amssymb,amsthm,amstext,amscd,eucal,srcltx}
\usepackage{epsfig,graphicx,bm}
\usepackage{epstopdf, epsf}
\usepackage{dcolumn}
\usepackage{comment}
\usepackage{hyperref}
\usepackage{tikz}
\usepackage[nointegrals]{wasysym}
\usepackage{braket}
\usepackage{etoolbox}
\DeclareMathOperator{\tr}{tr}

\newcommand{\ee}{\end{equation}}

\newcommand{\bc}{\begin{center}}
\newcommand{\ec}{\end{center}}

\usepackage{amsmath}
\usepackage{amssymb}
\usepackage{amsthm}
\usepackage{graphicx}
\usepackage{lineno}
\usepackage{float}
\usepackage[all]{xy}
\usepackage{mathrsfs}
\usepackage{tikz-cd}

\usepackage{xcolor}
\usepackage{tikz}
\usetikzlibrary{tqft}

\newcommand{\bea}{\begin{eqnarray}}
\newcommand{\eea}{\end{eqnarray}}

\usepackage{braket}

\newcommand{\Tr}{{\rm Tr}}

\hypersetup{hidelinks}

\begin{document}

\title{Correlation Geometry and Emergent Space-Time from Stochastic Correlation Flows}

\author{Antonino Marcian\`o}
\email{marciano@fudan.edu.cn}
\affiliation{Center for Field Theory and Particle Physics \& Department of Physics, Fudan University, 200433 Shanghai, China}
\affiliation{Laboratori Nazionali di Frascati INFN, Frascati (Rome), Italy, EU}
\affiliation{{INFN, Sezione di Roma Tor Vergata, I-00133 Rome, Italy}}

\begin{abstract}
\noindent
\begingroup
We propose an effective framework relating correlation geometry to stochastic flows of correlation projectors. Physical configurations are encoded by measures on spaces of self-adjoint operators, with physical equivalence defined by unitary conjugation. Assuming a smooth, spectrally separated correlation sector, local orthonormal frames define a Wilczek--Zee connection, while a Stratonovich gradient-flow ansatz preserves the projector constraints. In a local continuum approximation with massive fast modes, standard one-loop methods generate Yang--Mills-type terms and curvature terms of Einstein--Hilbert form. Along a smooth closed path, the trace of a cyclic product of projectors converges to the trace of the parallel-transport operator as the discretization is refined. We examine possible geometric, disordered, and topological regimes and motivate the stochastic dynamics through an effective open-system description. The first-order frame action defines a symplectic structure on the projector sector, providing a starting point for its relation to gravitational canonical variables. We identify the matching conditions required for an Ashtekar phase-space description and the assumptions connecting correlation geometry to induced interactions and emergent spacetime.
\endgroup

\end{abstract}

\maketitle

\section{Introduction}
\noindent 
Understanding the emergence of spacetime and gauge interactions from more fundamental quantum structures remains one of the central challenges of theoretical physics. A growing body of work suggests that geometry itself may not be fundamental, but rather arise from collective organization of microscopic quantum degrees of freedom. Notable examples include emergent gravity in group field theory and tensor models \cite{Liu:2023gft}, relatively local Hamiltonians \cite{Lee:2018A,Lee:2018B,Lee:2019wqb}, and causal fermion systems \cite{Finster:2016}.

A common feature of these approaches is that geometry and locality are encoded in patterns of correlations or entanglement. In particular, it has been argued that spacetime dimension, topology, and metric structure may be reconstructed from collective variables defined on entanglement networks \cite{Liu:2023gft}, while gauge symmetries can emerge as redundancies in the description of these collective degrees of freedom \cite{Lee:2018A,Lee:2018B}.

In this work we develop a framework in which these ideas are formulated in terms of \emph{correlation geometry}, where physical configurations are described by measures on nonlinear spaces of operators \cite{Paganini:2025}. In this setting, gauge invariance is realized as unitary equivalence, and geometric structures arise as effective descriptions of organized correlation sectors.

{We propose a phenomenological stochastic gradient flow for coarse-grained correlation sectors. This is an additional effective model, not a derivation from or replacement for the causal action principle.} This flow is motivated by both stochastic quantization \cite{Parisi:1981} and by coarse-graining in many-body quantum systems \cite{Kadanoff:1975,TarjusTissier:2004,TarjusTissier:2006,TarjusTissier:2011,Migdal:1976}, where effective dynamics generically acquires stochastic contributions due to unresolved degrees of freedom.

Within this approach, local correlation subspaces define moving frames in Hilbert space, and their transport induces a non-Abelian Berry connection \cite{WilczekZee:1984,Berry:1984}. {Given a local continuum regime and a suitable massive fluctuation operator, this connection supports an induced gauge action. Standard heat-kernel methods then generate Yang--Mills and curvature terms \cite{Sakharov:1968,Vassilevich:2003}; the continuum metric and fluctuation spectrum remain inputs to this calculation.}

{Furthermore, we examine the frame symplectic structure and the additional matching conditions needed to relate it to the Ashtekar phase space \cite{Ashtekar:1991}.} Gauge-invariant observables are naturally expressed in terms of correlation projectors, providing a direct interpretation of Wilson loops as cyclic overlap observables.

{The framework suggests candidate geometric, disordered, and topological regimes. Establishing distinct phases requires a specified action, observables, and an appropriate large-system limit; identifying stochastic relaxation with renormalization-group evolution additionally requires a coarse-graining prescription.}

{The construction provides a common effective language for investigating gauge transport, induced curvature terms, and possible phase organization in correlation geometry.}

\section{Correlation Geometry and Stochastic Gauge--Geometry Flow}
\noindent 
Correlation geometry provides a kinematical framework in which physical configurations are described by a triple
\begin{equation}
(H, \mathcal{F}_{p,q}, \rho),
\end{equation}
{where $H$ is a separable complex Hilbert space, $\mathcal{F}_{p,q}$ consists of finite-rank self-adjoint operators with at most $p$ positive and $q$ negative eigenvalues, and $\rho$ is a positive Borel measure. Fixed-rank, fixed-signature subsets form its regular strata \cite{Paganini:2025}.} Physical equivalence is defined in terms of unitary transformations acting as
\begin{equation}
x \;\to\; U x U^{-1},
\end{equation}
under which the measure transforms accordingly.

The space of physical configurations is therefore given by equivalence classes of measures under unitary conjugation, establishing a direct analogy with gauge orbit spaces in Yang--Mills theory.

The presence of a variational principle on the space of measures naturally suggests a dynamical extension. We promote the measure to a time-dependent object,
\begin{equation}
\rho \;\to\; \rho_t,
\end{equation}
and consider its evolution governed by a stochastic gradient flow,
\begin{equation}
d\rho_t = - \nabla_{\rho} \mathcal{S}[\rho_t]\, dt + \mathcal{N}[\rho_t],
\end{equation}
{where $\mathcal{S}[\rho]$ is an effective functional and $\mathcal{N}$ denotes a stochastic increment. This measure-level equation is schematic: its metric, noise covariance, and positivity and normalization constraints must be specified to define a process. Below we use a finite-dimensional, fixed-rank projector model instead. The flow parameter is auxiliary and is not identified with physical time.}\\ 

To extract effective geometric information from the correlation sector, we associate to the measure $\rho_t$ a coarse-grained correlation operator
\begin{equation}
\mathcal{C}[\rho_t] = \int_{\mathcal{F}_{p,q}} x \, d\rho_t(x)\,,
\end{equation}
{assuming a finite first operator moment (automatically well defined with a bounded finite-dimensional regulator), or, more generally, a suitable correlation kernel constructed from moments of the measure.} The slowly varying sector of the theory is then identified through a spectral projection
\begin{equation}
P_t = \chi_{\Lambda(t)}\!\left(\mathcal{C}[\rho_t]\right),
\end{equation}
{where $\chi_{\Lambda}$ selects a spectral interval separated from its complement by a nonzero gap. Calling this sector low energy requires an additional relation between the correlation operator and physical energies. A local field $P(x,\tau)$ further requires local correlation data; it is not reconstructed from this single global first moment alone.}

The image of $P_t$ defines a time-dependent correlation subspace
\begin{equation}
S_t = \mathrm{Im}(P_t) \subset \mathcal{H}\,.
\end{equation}
In this way, the stochastic evolution of the measure $\rho_t$ induces a dynamical trajectory
\begin{equation}
\rho_t \quad \longrightarrow \quad P_t \quad \longrightarrow \quad S_t
\end{equation}
in the Grassmannian of subspaces of $\mathcal{H}$.\\

A choice of local frame for $S_t$ defines a representative of the corresponding unitary orbit. The transport of this frame along the stochastic trajectory induces a natural connection on the bundle of correlation subspaces, which will be identified below with the emergent gauge field.

Since the evolving projector $P_t$ defines a trajectory in the Grassmannian of subspaces of $\mathcal H$, the corresponding correlation subspace
\begin{equation}
S_t = \mathrm{Im}(P_t)
\end{equation}
changes continuously along the stochastic flow. To compare nearby subspaces, one must choose a local frame
\begin{equation}
\psi_t \in S_t\,,
\end{equation}
which is defined only up to a time-dependent phase (or, more generally, a local unitary rotation in the higher-rank case).

The ordinary derivative $\partial_t \psi_t$ does not in general remain tangent to the chosen horizontal subspace, and therefore comparison of nearby frames requires a connection. The natural connection induced by the transport of correlation frames is
\begin{equation}
A_t = i \langle \psi_t | \partial_t \psi_t \rangle\,,
\end{equation}
which measures the vertical component of frame transport along the trajectory in projector space.\\

Equivalently, the covariant derivative
\begin{equation}
{D_t \psi_t = \partial_t \psi_t + i A_t \psi_t}
\end{equation}
defines horizontal transport of the local correlation frame. Under a local phase redefinition
\begin{equation}
\psi_t \rightarrow e^{i\chi(t)} \psi_t\,,
\end{equation}
the connection transforms as
\begin{equation}
{A_t \rightarrow A_t - \partial_t \chi}\,,
\end{equation}
showing that $A_t$ is the natural $U(1)$ connection associated with the bundle of correlation subspaces.\\

For higher-rank projectors, the same construction yields the non-Abelian Wilczek--Zee connection associated with transport on the Grassmannian of correlation sectors. {The connection defines parallel transport; nonzero curvature or holonomy alone does not imply a topologically nontrivial bundle. For stochastic trajectories the transport is understood in the Stratonovich sense.}\\

The physical content of the theory is encoded in the support of the measure $\mathrm{supp}(\rho) \subset \mathcal{F}_{p,q}$. Under stochastic evolution, this support may undergo nontrivial deformations, including fragmentation and concentration. {They may alter the topology of the measure support, rather than that of the ambient operator space; no such change is established by the fixed-rank projector flow alone.}

Gauge-invariant observables analogous to Wilson loops probe the global organization of the support of the measure
\begin{equation}
\mathrm{supp}(\rho_t) \subset \mathcal{F}_{p,q}\,.
\end{equation}
Under stochastic evolution, the support of $\rho_t$ may become concentrated on distinguished lower-dimensional sectors of the operator manifold $\mathcal{F}_{p,q}$, such as specific unitary orbits, degenerate spectral strata, or topologically nontrivial sub-manifolds.

In the infrared description, transport of correlation frames becomes effectively constrained to these sectors. This can induce nontrivial holonomy and long-range disordering effects in correlation-loop observables, suggesting a possible mechanism for confinement analogous to the dominance of topological sectors in non-Abelian gauge theories.\\

{Configurations that do not admit a stable smooth continuum description will be called pre-geometric here; being out of equilibrium alone is not sufficient for this classification.}\\

In this framework, gauge invariance, geometry, and dynamics are unified within a single structure $(H, \mathcal{F}_{p,q}, \rho_t)$. {Gauge transport is encoded by correlation frames; a smooth infrared geometric regime is assumed in the effective calculations that follow.}

\section{Induced Gauge Dynamics from Stochastic Correlation Flow}
\noindent
For $H=\mathbb{C}^N$, let us consider a rank-$k$ projector
\begin{equation}
P^2=P, \qquad P^\dagger=P, \qquad \Tr P = k.
\end{equation}
Equivalently, one may introduce an orthonormal frame
\begin{equation}
\Psi = \big(\psi^1,\dots,\psi^k\big), \qquad \Psi^\dagger \Psi = \mathbb{I}_k,
\end{equation}
such that
\begin{equation}
P=\Psi\Psi^\dagger.
\end{equation}

The representation of the projector in terms of frames is not unique. Indeed,
\begin{equation}
P = \Psi \Psi^\dagger
\end{equation}
is invariant under unitary frame rotations, namely
\begin{equation}
\Psi \rightarrow \Psi V, \qquad V \in U(k)\,.
\end{equation}
At the purely kinematical level, this is a redundancy in the choice of basis for the correlation subspace $\mathrm{Im}(P)$.

Local gauge structure emerges once the coarse-grained correlation sector admits an effective continuum description in terms of slowly varying projectors
\begin{equation}
P(x)\,,
\end{equation}
which define local correlation subspaces over the emergent manifold. In this regime, neighboring spacetime regions correspond to nearby sectors in correlation geometry, and the frame redundancy acts independently on each local subspace,
\begin{equation}
\Psi(x) \rightarrow \Psi(x) V(x)\,.
\end{equation}

The resulting structure is therefore a local $U(k)$ bundle over the emergent spacetime, with the associated connection arising from transport of local correlation frames.\\

The associated Wilczek--Zee connection is
\begin{equation} \label{WZ}
A_\alpha = i\Psi^\dagger \partial_\alpha \Psi ,
\end{equation}
where $\alpha$ denotes either the stochastic flow parameter $\tau$ or an emergent spacetime coordinate $x^\mu$.

This expression should be understood in the geometric, coarse-grained regime in which the support of the measure
\begin{equation}
\mathrm{supp}(\rho_\tau)\subset \mathcal F_{p,q}
\end{equation}
admits a smooth local parametrization by emergent coordinates $x$. In that regime the correlation sector is described by a smooth projector field
\begin{equation}\label{pepsi}
P(x,\tau)=\Psi(x,\tau)\Psi^\dagger(x,\tau),
\end{equation}
and the image of $P(x,\tau)$ defines a local correlation subspace of $\mathcal H$.

The local unitary transformation
\begin{equation}
\Psi(x,\tau)\rightarrow \Psi(x,\tau)V(x,\tau),
\qquad V(x,\tau)\in U(k),
\end{equation}
is therefore not an arbitrary unitary transformation on the full operator space $\mathcal F_{p,q}$. It is a smooth change of frame inside the fiber
\begin{equation}
\mathrm{Im}\,P(x,\tau)\subset \mathcal H
\end{equation}
over each point of the emergent correlation geometry.\\

Similarly, $V(x,\tau)$ is determined locally by the freedom to choose different orthonormal bases for the same projector field $P(x,\tau)$. The projector is invariant,
\begin{equation}
P(x,\tau)\rightarrow P(x,\tau),
\end{equation}
while the connection transforms as
\begin{equation}
A_\alpha \rightarrow V^{-1}A_\alpha V+iV^{-1}\partial_\alpha V .
\end{equation}

Thus the gauge connection is the connection on the bundle of local frames over the smooth family of correlation projectors. Its existence depends on the emergence of a smooth local description of $\mathrm{supp}(\rho_\tau)$ in $\mathcal F_{p,q}$; outside such a geometric phase, one should not assume a conventional local gauge field description.\\

We endow the projector with a stochastic gradient flow in an auxiliary time $\tau$,
\begin{equation}
dP_\tau
=
-\,\Pi_{P_\tau}\!\left(\frac{\delta \mathcal{S}[P]}{\delta P}\right)d\tau
+
\sqrt{2D}\,\Pi_{P_\tau}\circ dW_\tau,
\end{equation}
where $D$ is a diffusion coefficient, $dW_\tau$ is a matrix-valued Wiener increment, and $\Pi_P$ denotes the projection onto the tangent space of the Grassmannian. A convenient choice is
\begin{equation}
\Pi_P(X)=[P,[P,X]],
\end{equation}
{For Hermitian gradients and Hermitian matrix noise, this tangent projection preserves the projector constraints in the Stratonovich equation. We use the Hilbert--Schmidt metric and a finite spatial regulator when treating projector fields.}\\

In the geometric infrared regime, the slowly varying correlation sector is described by the projector field in Eq.~\eqref{pepsi}, 
which is defined 
over an emergent spacetime manifold $\mathcal M$. The corresponding frame field $\Psi(x,\tau)$ hence defines a local correlation basis, and the associated Wilczek--Zee connection \eqref{WZ}. 
At the microscopic level, the connection is therefore a composite quantity completely determined by the local correlation frame. The fundamental dynamical variable remains the projector sector itself.\\

{As an effective ansatz, we use the following Euclidean action on each fixed-$\tau$ slice. In this action $\alpha$ runs over the $d$ continuum coordinates only; $\tau$ labels the stochastic evolution and is not an additional coordinate integrated in the quantum determinant. The action is}
\begin{eqnarray}
S[\Psi,A]
=
&&\frac{\kappa}{2}
{\int d^dx\,}
{\rm Tr}
\Big[
(D_\alpha\Psi)^\dagger
(D_\alpha\Psi)
\Big]
+ \nonumber\\
&& \phantom{a} {\int d^dx\,} V(P)\,,
\end{eqnarray}
with
\begin{equation}
{D_\alpha\Psi
=
\partial_\alpha\Psi
+
i\Psi A_\alpha \,.}
\end{equation}

The potential term $V(P)$ should be understood as the local infrared approximation to the underlying correlation action governing the measure $\rho$ on $\mathcal F_{p,q}$. {Its microscopic origin is not derived here. On a fixed-rank Grassmannian a single-point potential invariant under all ambient unitary conjugations is constant; a nontrivial $V$ therefore requires specified background data or couplings between correlation sectors.}\\

Although $A_\alpha$ is microscopically determined by $\Psi$, coarse-graining separates slow collective modes from fast frame fluctuations. {Treating the connection as an autonomous collective field is an additional effective-field assumption. A constrained integration or an auxiliary-field construction is required to establish it for a microscopic projector model; Appendix~\ref{app:oneloop} illustrates an auxiliary-field calculation.} This is analogous to the emergence of collective hydrodynamic or gauge variables in effective field theory. In this sense, the local field theory for $(P,\Psi,A)$ should be viewed as a collective effective description of the underlying nonlocal correlation dynamics on $\mathcal F_{p,q}$.\\

The effective dynamics of the emergent connection is obtained by coarse-graining the fast fluctuations of the correlation sector. Since the physical configuration is encoded in the projector $P=\Psi\Psi^\dagger$, the frame field $\Psi$ should be regarded as a local parametrization of the projector degrees of freedom, rather than as an independent physical variable. The redundancy $\Psi \sim \Psi V$ then corresponds to vertical directions in the frame bundle and does not represent physical fluctuations. The functional integration therefore concerns horizontal deformations of the correlation sector, locally represented by fluctuations of $\Psi$ modulo frame redundancy.\\

Introducing a slowly varying background connection $A_\alpha$, one obtains the coarse-grained effective action
\begin{equation}
e^{-S_{\rm eff}[A]}
=
\int \mathcal D\Psi_\perp\,
e^{-S[\Psi,A]},
\end{equation}
where $\Psi_\perp$ denotes the horizontal fluctuation modes associated with deformations of the projector sector.\\

Similarly, this may be viewed as a background-field expansion around slowly varying correlation configurations, in which the emergent connection captures the transport properties of the infrared correlation structure.

Expanding around a slowly varying background then yields, at one-loop order,
\begin{equation}
S_{\rm eff}[A]
=
\frac12
{\rm Tr}
\log
\left(
-D_\alpha D^\alpha
+
M^2
+
\cdots
\right).
\end{equation}
The functional integral therefore does not sum over arbitrary choices of basis, but over inequivalent deformations of the correlation geometry itself.\\

Expanding the frame field around a slowly varying background,
\begin{equation}
\Psi=\Psi_0+\varphi ,
\end{equation}
and retaining only the quadratic part of the action for the fast horizontal fluctuations $\varphi$, one obtains schematically
\begin{equation}
S[\Psi,A]
=
S[\Psi_0,A]
+
\frac12
{\int d^dx\,}
\varphi^\dagger
\Delta_A
\varphi
+\cdots ,
\end{equation}
where $\Delta_A$ is the fluctuation operator in the background connection $A_\alpha$.

{Assuming a stable background, separation of fast massive modes, and a Laplace-type quadratic operator, we write schematically}
\begin{equation}
\Delta_A
=
-D_\alpha D^\alpha+M^2+\mathcal{R}[P,A]+\cdots ,
\end{equation}
{where $D_\alpha$ is the induced covariant derivative in the fluctuation representation (for column coefficients transforming as $c\to V^{-1}c$, $D_\alpha c=(\partial_\alpha-iA_\alpha)c$),} $M$ denotes the mass scale of the fast correlation-sector fluctuations, and $\mathcal{R}[P,A]$ stands for possible curvature, projector-dependent, or interaction terms.

{Using real coordinates for the independent fast fluctuation modes, their Gaussian integral gives}
\begin{equation}
e^{-S_{\rm eff}[A]}
\propto
\left(\det \Delta_A\right)^{-1/2},
\end{equation}
and therefore
\begin{equation}
S_{\rm eff}[A]
=
\frac12
{\rm Tr}\log \Delta_A .
\end{equation}

Thus the expression
\begin{equation}
S_{\rm eff}[A]
=
\frac12
{\rm Tr}\log
\left(
-D_\alpha D^\alpha+M^2+\cdots
\right)
\end{equation}
is the one-loop determinant of the fast horizontal projector/frame fluctuations in the slowly varying background connection. The trace includes both integration over spacetime and summation over internal frame indices.\\

The derivative expansion of the functional determinant yields a local gauge-invariant effective action of the form
\begin{equation}
S_{\rm eff}[A]
=
\int d^dx \left[
\frac{1}{4g_{\rm eff}^2}F_{\mu\nu}^aF^{a\mu\nu}
+
\mathcal{O}(F^3,\partial F \partial F)
\right],
\end{equation}
where
\begin{equation}
F_{\mu\nu}
=
\partial_\mu A_\nu-\partial_\nu A_\mu-i[A_\mu,A_\nu].
\end{equation}
{Thus a Yang--Mills-type term is induced under the stated local fluctuation assumptions; establishing an autonomous Yang--Mills theory from the original measure requires further microscopic matching.}\\

{For the deterministic limit and the composite connection, $\Tr[(D_\mu\Psi)^\dagger D_\mu\Psi]=\tfrac12\Tr(\partial_\mu P\,\partial_\mu P)$. The stated kinetic normalization therefore gives}
\begin{equation}
\partial_\tau P
=
-\, [P,[P,\mathcal{H}_{\rm eff}[P]]],
\end{equation}
{with the gradient generator (not the unitary Hamiltonian of Appendix~\ref{app:lindblad})}
\begin{equation}
\mathcal{H}_{\rm eff}[P]
=
{-\frac{\kappa}{2}\,\partial_\mu\partial^\mu P}
+
\frac{\partial V(P)}{\partial P}.
\end{equation}
Fixed points satisfy
\begin{equation}
[P,[P,\mathcal{H}_{\rm eff}[P]]]=0.
\end{equation}
This condition implies that the effective Hamiltonian preserves the subspace selected by $P$, so that the corresponding orbit of frames is stationary under coarse-graining.

{Frame rotations are gauge redundancy and do not constitute physical gapless modes; fixed-rank projectors have no radial modes. If a suitable extension has massive physical modes and a separate slow collective connection, integrating out the massive modes can yield the leading term}
\begin{equation}
S_{\rm IR}[A]
=
\frac{1}{4g_*^2}\int d^dx\, F_{\mu\nu}^aF^{a\mu\nu}
+\cdots.
\end{equation}
{Whether a specified stochastic model reaches a Yang--Mills universality class is an open dynamical question. A stationary projector configuration is not by itself a renormalization-group fixed point.}\\

The resulting picture is that gauge invariance is encoded kinematically in the unitary orbit structure of the correlation frame, while gauge dynamics is generated dynamically by stochastic coarse-graining. {Stable stationary sectors are candidates for an infrared description with induced gauge terms, subject to the separation and matching assumptions above.}

\section{Phase structure and topological sectors}
\noindent
In the emergent-geometry framework of Ref.~\cite{Liu:2023gft}, spacetime structure arises from the entanglement of underlying quantum degrees of freedom. In the present formulation, this structure is naturally encoded in terms of correlation geometry, where local physical information is captured by a projector
\begin{equation}
P(x,\tau)=\Psi(x,\tau)\Psi^\dagger(x,\tau),
\qquad
\Psi^\dagger\Psi=\mathbb I_k,
\end{equation}
onto a coarse-grained correlation subspace.

As discussed above, slowly varying correlation projectors define local correlation subspaces together with an emergent $U(k)$ connection describing the transport of correlation frames. In the geometric infrared regime, the corresponding gauge structure provides the natural variables for the effective canonical description.\\

\begingroup
To examine a possible canonical description, we separately postulate a first-order frame action,
\begin{equation}
S_{\rm frame}=\int dt\int_\Sigma d^3x\,
\left[i\kappa\Tr(\Psi^\dagger\partial_t\Psi)-{\cal H}_{\rm frame}\right],
\end{equation}
with a frame-invariant Hamiltonian functional ${\cal H}_{\rm frame}$, left unspecified here. This is additional input, not a consequence of the preceding Euclidean gradient-flow action. Locally, time-dependent frame rotations change its kinetic term by a boundary term. The ordinary derivative is intentional: substituting the composite connection into the horizontal derivative would give $\Psi^\dagger D_t\Psi=0$.
The associated local symplectic potential is
\begin{equation}
\Theta_{\rm frame}=i\kappa\int_\Sigma d^3x\,
\Tr(\Psi^\dagger\delta\Psi).
\end{equation}
Its exterior derivative descends to the Grassmannian after quotienting frame rotations, as discussed in Appendix~\ref{app:symplectic}. For $k=2$ and $\tau_i=\sigma_i/2$, the traceless connection components are
\begin{equation}
A_a^i=2\Tr(\tau_i A_a)
=2i\Tr(\tau_i\Psi^\dagger\partial_a\Psi).
\end{equation}
If a local family of frames $\Psi[A]$ can be chosen, its pullback can be written formally as
\begin{equation}
\Theta_{\rm red}[A]=\int_\Sigma d^3x\,
E_i^a[A],\delta A_a^i.
\end{equation}
Here $E[A]$ is a functional coefficient, not an independently established densitized triad. Consequently this expression does not imply canonical brackets for independent $(A,E)$ fields. A nondegenerate triad, an appropriate phase-space map, and the gravitational constraints must be supplied before identifying the reduced system with canonical gravity. Appendix~\ref{app:pullback} makes this limitation explicit.
\endgroup

Integrating out fast fluctuations of the correlation frame generates an effective action for the emergent fields. {Under the continuum and fluctuation assumptions above, the one-loop derivative expansion permits both Yang--Mills and curvature terms,}
\begin{equation}
S_{\rm eff}
\!=\!
\int \!\! d^4x \sqrt{g}
\left[
\Lambda_{\rm ind}
\!+\!
{c_R R}
\!+\!
\frac{1}{4g_{\rm ind}^2}F_{\mu\nu}^aF^{a\mu\nu}
\!+\!\cdots
\right].
\end{equation}

\begingroup
The curvature term is of Einstein--Hilbert form, but its coefficient $c_R$ depends on the field content, regulator, and curvature couplings (Appendix~\ref{app:EH}). The determinant assumes a Riemannian continuum metric; it does not reconstruct that metric or establish a Lorentzian continuation.

If a Lorentzian gravitational sector with a nondegenerate triad is obtained independently, it can be written in real Ashtekar--Barbero variables \cite{Ashtekar:1991,Barbero:1995},
\begin{equation}
\mathcal A_a^i=\Gamma_a^i[e]+\gamma K_a^i,
\qquad
E_i^a=\tfrac12\epsilon^{abc}\epsilon_{ijk}e_b^{\ j}e_c^{\ k},
\end{equation}
with spin connection $\Gamma[e]$, extrinsic curvature $K$, and real nonzero Barbero--Immirzi parameter $\gamma$. Their gravitational symplectic form is
\begin{equation}
\Omega_{\rm GR}=\frac{1}{8\pi G\gamma}
\int_\Sigma d^3x\,\delta E_i^a\wedge\delta\mathcal A_a^i.
\end{equation}
This is a target for matching, not a result of the frame pullback. Identifying the Berry connection $A$ with $\mathcal A$ and recovering the Gauss, diffeomorphism, and Hamiltonian constraints remain to be demonstrated.
\endgroup

Gauge-invariant observables admit a natural interpretation in terms of transport of correlation subspaces. Consider a closed path $C$ in the emergent infrared geometry, discretized by points
\begin{equation}
x_1,x_2,\ldots,x_N,
\qquad
x_{N+1}=x_1.
\end{equation}

Each projector
\begin{equation}
P(x_n)
\end{equation}
defines a local correlation subspace in $\mathcal H$. Neighboring projectors generally have nontrivial overlap, so that successive products
\begin{equation}
P(x_{n+1})P(x_n)
\end{equation}
describe the local transport of correlation states between nearby sectors.

The cyclic product
\begin{equation}
W_P(C)
=
{\rm Tr}
\Big(
{P(x_N)\cdots P(x_2)P(x_1)}
\Big)
\end{equation}
therefore measures the cumulative overlap acquired along the closed path $C$.

{For a smooth, fixed-rank projector along $C$, the reverse-ordered product above converges under mesh refinement to the trace of forward parallel transport. At finite mesh the overlaps need not be unitary. With this orientation and the convention $A=i\Psi^\dagger d\Psi$, the limit is}
\begin{equation}
W(C)
=
{\rm Tr}\,
{\cal P}
\exp
\left(
i\oint_C A_\mu dx^\mu
\right).
\end{equation}

In this sense, Wilson loops admit a correlation-theoretic interpretation as observables probing the transport and twisting of local correlation subspaces along closed paths in the emergent geometry. {Here $\mathcal P$ denotes path ordering, whereas $P(x)$ denotes a projector. Reversing the ordering reverses the loop orientation.}\\

This structure is closely analogous to the role played by operator products in causal fermion {systems~\cite{Finster:2016}}, where geometric and causal information is encoded directly in spectral properties of compositions of correlation operators. {Both frameworks organize geometric information through operators; the continuum metric used in our effective calculation, however, is assumed rather than reconstructed.}

From this perspective, the cyclic overlap observable
\begin{equation}
{\rm Tr}
\Big(
P(x_1)\cdots P(x_N)
\Big)
\end{equation}
{is a transport observable. It is not the causal action, which is a variational functional of the measure with a specified spectral Lagrangian.} The Wilson holonomy then emerges as the continuum infrared limit of these discrete correlation observables. The present framework may therefore be viewed as extending operator-based geometric constructions toward a dynamical transport theory for correlation sectors, in which gauge holonomy arises from collective organization of projector overlaps.\\

\begingroup
Nontrivial holonomy is a transport diagnostic, not by itself a criterion for a topological defect or a nontrivial bundle. For rank $k$, the trivial unnormalized Wilson trace is $W(C)=k$, rather than $1$. A topological sector requires an appropriate homotopy class or characteristic number and boundary conditions.

For example, in the $CP^1$ prototype,
\begin{equation}
P=\tfrac12(\mathbb I+n^i\sigma_i),\qquad n^in^i=1,
\end{equation}
the convention $A=i\psi^\dagger d\psi$ gives
\begin{equation}
f_{ab}=-\tfrac12 n\cdot(\partial_a n\times\partial_b n).
\end{equation}
Since $CP^1\simeq S^2$ has $\pi_1=0$, it does not support protected loop-winding vortices by itself. On a compactified oriented two-dimensional surface, smooth textures can instead carry the integer degree associated with $\pi_2(S^2)=\mathbb Z$, measured, up to orientation, by $(2\pi)^{-1}\int f$. Vortex sectors would require additional order-parameter or boundary structure.
\endgroup

{We use the following descriptions for candidate infrared regimes; no phase diagram or phase transition is computed here.}

In a \emph{geometric phase}, neighboring projectors vary slowly across the emergent infrared manifold,
\begin{equation}
P(x+\delta x)-P(x)\ll 1 ,
\end{equation}
so that local correlation subspaces admit coherent transport and an effective continuum spacetime description. In this regime, the induced connection and metric variables provide a stable long-wavelength geometric description.

In a \emph{pre-geometric phase}, the correlation sector fluctuates strongly under the stochastic flow, and no stable hierarchy of slowly varying projector configurations emerges. Correlation transport becomes incoherent, preventing the construction of a persistent infrared geometry or local effective spacetime structure.

In a \emph{topological phase}, long-range transport is dominated by nontrivial holonomy sectors. Although local geometric order may be partially disrupted, global transport observables such as correlation-loop overlaps remain stable and are controlled by topological winding of the projector sector in operator space.

These phases are therefore distinguished by the large-scale behavior of projector correlations, the stability of infrared transport structures, and the organization of holonomy under coarse-graining. From the viewpoint of the stochastic flow, different phases correspond to distinct infrared attractors of the coarse-grained correlation dynamics.\\

{A genuine phase transition would require nonanalytic observables in a specified thermodynamic or continuum limit. On a fixed-rank Grassmannian, $P$ has only eigenvalues $0$ and $1$: a gap closing refers instead to the underlying correlation or fluctuation operator. Rank changes require leaving the smooth fixed-rank description.}\\

The coarse-grained evolution of the projector sector is constrained by the requirement that the projector condition
\begin{equation}
P^2=P
\end{equation}
be preserved under the flow. The natural dissipative evolution on the Grassmannian of projectors is therefore obtained by projecting the effective gradient force onto the tangent space of the projector manifold.

This yields the deterministic flow
\begin{equation}
\partial_\tau P
=
-[P,[P,H_{\rm eff}[P]]] ,
\end{equation}
where the double commutator implements the tangent projection and guarantees preservation of the projector constraint throughout the evolution.

Physically, $H_{\rm eff}[P]$ represents the effective coarse-grained generator governing the slow correlation sector.

In addition, tracing out unresolved microscopic degrees of freedom induces stochastic fluctuations in the reduced dynamics. The resulting mesoscopic evolution therefore acquires a noise contribution,
\begin{equation}
\partial_\tau P
=
-[P,[P,H_{\rm eff}[P]]]
+
\eta ,
\end{equation}
where $\eta$ denotes the effective stochastic force generated by the microscopic environment.

{Appendix~\ref{app:lindblad} gives an open-system motivation for this effective ansatz. A general Lindblad generator does not imply this gradient drift: a closed projector description, relaxational dynamics, and the noise covariance are additional assumptions.}\\

{The auxiliary parameter $\tau$ may be related to a coarse-graining scale only after a scale map and effective-action flow are specified. Stochastic relaxation and Wilsonian renormalization are not identified by the equations given here.}

In this way, the framework unifies emergent gauge structure, gravitational dynamics, and phase transitions within a single correlation-geometric description.\\

\begingroup
The topological interpretation must therefore be tied to specified boundary conditions and invariant charges. Smooth evolution within a fixed-rank sector preserves a texture's homotopy class unless regularity or the spectral-gap assumption fails.
\endgroup

This perspective also suggests a natural notion of phase transition between correlation geometries. Since the fundamental variables are projectors, frames, and measures on operator space, distinct phases are characterized by qualitative changes in the organization of the coarse-grained correlation sector. {Candidate changes include closure of a correlation-operator gap, loss of the fixed-rank approximation, reorganization of the measure support, or altered large-distance loop correlations. These possibilities are not consequences of the finite-dimensional fixed-rank flow alone.}\\

An important feature of the topological regime is that long-distance physics becomes controlled by nontrivial holonomy sectors of the correlation transport structure. In this case, confinement properties may be characterized through the asymptotic behavior of correlation-loop observables.

In particular, the cyclic projector overlaps introduced above provide a correlation-theoretic analogue of Wilson-loop diagnostics. {In a specified gauge theory and probe representation, area-law behavior of the ensemble-averaged, normalized continuum Wilson loop can diagnose confinement. No such scaling is calculated here; finite-mesh overlap attenuation and screening by dynamical matter must be distinguished from a confining area law.}

From the viewpoint of correlation geometry, confinement is therefore associated with large-scale disordering of correlation transport rather than with fundamental localization in spacetime itself.\\

The stochastic correlation flow
\begin{equation}
\partial_\tau P
=
-[P,[P,H_{\rm eff}[P]]]
+
\eta
\end{equation}
therefore acquires the interpretation of a non-equilibrium coarse-graining dynamics on the space of correlation sectors.

Different infrared phases correspond to distinct long-wavelength attractors of this stochastic evolution. In particular, the large-scale organization of projector correlations, transport coherence, and holonomy structure may change qualitatively under the flow.

{Possible critical behavior would involve a correlation-operator gap closing or nonanalytic long-distance observables in a suitable limit, rather than a gap closing in the spectrum of a fixed-rank projector.}

In this sense, phase transitions correspond not to transitions between pre-existing spacetime manifolds, but to transitions between different emergent geometric regimes associated with inequivalent infrared organizations of the correlation sector. {Their possible dependence on coarse-graining remains a question for a specified microscopic model and scale map.}\\

\section{Emergent gravity}
\noindent 
{The present construction is motivated by the emergent-gravity mechanism developed in Ref.~\cite{Liu:2023gft},} where the entanglement structure of GFT graph states, together with a scalar relational clock, generates an effective Hamiltonian whose continuum limit reproduces gravitational dynamics and whose evolution can be interpreted as an RG flow related to a simplified Ricci flow. In that framework, the collective variables extracted from the entanglement pattern determine the emergent dimension, topology, and geometry. This logic is closely aligned with Lee's relatively local Hamiltonians {\cite{Lee:2018B,Lee:2019wqb}}, in which locality and geometry are state dependent and are controlled by the entanglement structure of the microscopic state. \\

From this perspective, the coarse-grained entanglement sector may be represented by a moving projector onto a local correlation subspace of the Hilbert space. A choice of frame for this subspace induces a Wilczek--Zee connection, which provides the natural emergent gauge variable associated with frame transport, while stochastic coarse-graining of the RG evolution promotes the deterministic flow of collective variables to a stochastic gradient flow on the space of correlation sectors. In this way, the present formulation unifies Lee's state-dependent locality and the GFT entanglement-clock mechanism of Ref.~\cite{Liu:2023gft} within a common gauge-geometric language, and suggests that Yang--Mills dynamics arises as the infrared effective theory of fluctuating correlation frames.\\

The construction presented above admits a direct interpretation in terms of the emergent-gravity framework developed in Ref.~\cite{Liu:2023gft}, as well as in the context of relatively local Hamiltonians introduced by Lee {\cite{Lee:2018B,Lee:2019wqb}}. {We describe the proposed correspondence; an explicit map from the GFT state and its generator to the correlation measure and stochastic coefficients is not supplied.}

In Ref.~\cite{Liu:2023gft}, the fundamental degrees of freedom are encoded in Group Field Theory (GFT) graph states, whose entanglement structure determines the emergent spacetime properties. In particular, the Hilbert space is decomposed as a tensor product over sites,
\begin{equation}
\mathcal{H} = \bigotimes_{n} \mathcal{H}_n,
\end{equation}
and the entanglement graph state encodes adjacency and connectivity between sites.\\

The key result is that dimension, topology, and geometry emerge from collective variables constructed from this entanglement structure. These collective variables define an effective continuum description in which the gravitational Hamiltonian takes the form of the ADM Hamiltonian and can be recast in Ashtekar variables.\\

In the emergent-gravity framework of {Ref.~\cite{Liu:2023gft}}, locality and geometry arise from the entanglement structure of the underlying quantum state. The effective infrared degrees of freedom are associated with slowly varying collective sectors selected by the entanglement pattern.

In the present framework, this structure is reformulated in terms of correlation geometry. We associate to each coarse-grained region an effective correlation operator
\begin{equation}
F(x),
\end{equation}
whose matrix elements encode the local relational structure of the underlying quantum state.

The slowly varying infrared sector is then selected through a spectral projection
\begin{equation}
P(x,\tau)
=
\chi_{\Lambda(\tau)}
\!\left(
F(x)
\right),
\end{equation}
where $\chi_{\Lambda}$ denotes projection onto a low-energy or slowly varying spectral sector determined by the coarse-graining scale $\Lambda(\tau)$.

The image of $P(x,\tau)$ therefore defines the subspace of correlation modes that remain coherent under coarse-graining.

This construction differs conceptually from the tensor-product decomposition used in {Ref.~\cite{Liu:2023gft}}. There, locality is associated with entanglement among subsystem factors
\begin{equation}
\mathcal H=\bigotimes_n \mathcal H_n .
\end{equation}
{Here, by contrast, we model local correlation information using selected subspaces}
\begin{equation}
\mathrm{Im}(P(x,\tau))\subset\mathcal H
\end{equation}
inside a single underlying Hilbert space.

The projector sector may therefore be viewed as an operator-geometric generalization of the effective entanglement sectors of {Ref.~\cite{Liu:2023gft}}, where local structure emerges from coherence and transport properties of correlation subspaces rather than from a fixed tensor-product decomposition. In this sense, emergent locality is relational rather than subsystem-based.\\

A central ingredient of Ref.~\cite{Liu:2023gft} is the introduction of a scalar field $\phi$ serving as a relational clock. The evolution of the system is generated by an effective Hamiltonian derived from the GFT action, and the resulting dynamics can be interpreted as a renormalization-group flow, which in the continuum limit is related to a simplified Ricci flow.

{This structure motivates, but does not derive, the deterministic ansatz used here.} The coarse-graining scale $\Lambda(\tau)$ and the associated projector $P(x,\tau)$ evolve under a flow which, in the simplest approximation, can be written as
\begin{equation}
\partial_\tau P = -[P,[P,\mathcal{H}_{\rm eff}[P]]] + \eta,
\end{equation}
where $\mathcal{H}_{\rm eff}$ is an effective Hamiltonian functional and $\eta$ denotes stochastic contributions arising from integrating out microscopic degrees of freedom.

{Thus, the RG/Ricci flow of Ref.~\cite{Liu:2023gft} motivates a comparison with the effective stochastic model, without identifying the two evolution parameters.}

{A related stochastic-geometric mechanism is developed by Lulli, Marcian\`o, and Piscicchia~\cite{Lulli:2023srf}, who connect stochastic Ricci-flow dynamics to a GKLS-type master equation in a non-relativistic matter limit. This motivates relating geometric noise to open-system evolution; identifying the present projector flow with a Wilsonian RG flow still requires a coarse-graining map.}

At the microscopic level, the underlying correlation configuration is still governed by a variational principle, analogous to the minimizing dynamics of the causal action in causal fermion {systems~\cite{Finster:2016}}. The stochasticity does not replace this variational structure. {Rather, it is postulated as a mesoscopic model motivated by unresolved microscopic correlation modes.}

In this sense, the stochastic term encodes the effective fluctuations experienced by the infrared correlation sector when viewed as an open subsystem of the full operator dynamics. The deterministic part of the evolution still drives the system toward extrema of the effective correlation free-energy functional, while the noise term captures residual environmental fluctuations and non-equilibrium effects. {The stochastic flow is an effective thermodynamic ansatz compatible with a possible underlying variational description; deriving it from a causal action or a Wilsonian coarse-graining map is left open.}\\

The above picture is closely aligned with Lee's framework of relatively local Hamiltonians {\cite{Lee:2018B,Lee:2019wqb}}, where locality, geometry, and even dimensionality are state dependent and determined by the entanglement structure of the underlying quantum state.

In Lee's construction, the effective Hamiltonian depends on collective variables encoding the spatial metric, and generalized diffeomorphism invariance emerges from transformations acting on site labels. In Ref.~\cite{Liu:2023gft}, this structure is realized through shift and lapse tensors, whose algebra reproduces spatial and temporal diffeomorphisms.\\

\begingroup
For rank $k=2$, the frame construction has a $U(2)$ connection whose traceless part is locally $\mathfrak{su}(2)$ valued,
\begin{equation}
A_\mu=a_\mu\mathbb I_2+A_\mu^i\tau_i,
\qquad \tau_i=\sigma_i/2.
\end{equation}
A global reduction to an $SU(2)$ bundle requires a suitable trivialization of the determinant line. Neither this Lie-algebra decomposition nor the existence of a frame identifies $A$ with a gravitational connection. In particular, $\Psi^\dagger\partial_a\Psi$ transforms inhomogeneously and cannot by itself define a covariantly transforming triad. A separate triad construction and symplectic matching are required, as explained above.
\endgroup

The same structure gives a natural correlation-theoretic interpretation of Wilson loops. The holonomy of the emergent connection along a closed contour $C$ is
\begin{equation}
W(C)=\Tr\,{\cal P}\exp\!\left(i\oint_C A_\mu dx^\mu\right),
\end{equation}
which measures the parallel transport of the local correlation frame around $C$. Equivalently, in a discretized description one may define the cyclic overlap observable
\begin{equation}
{\cal W}(C)=\Tr\!\left({P(x_n)\cdots P(x_2)P(x_1)}\right),
\qquad x_{n+1}=x_1,
\end{equation}
whose continuum limit reproduces the Wilson holonomy of the induced connection. Wilson observables are therefore interpreted as closed-loop correlation observables probing the twisting of the local correlation sector.\\

Finally, the stochastic term in the projector flow admits a natural open-system interpretation in GFT. Writing the full Hilbert space as
\begin{equation}
{\cal H}_{\rm tot}={\cal H}_{\rm corr}\otimes{\cal H}_{\rm env},
\end{equation}
where ${\cal H}_{\rm corr}$ denotes the coarse-grained correlation sector and ${\cal H}_{\rm env}$ collects microscopic GFT modes that are not explicitly resolved, the reduced density matrix
\begin{equation}
\rho_{\rm corr}=\Tr_{\rm env}\rho_{\rm tot}
\end{equation}
obeys a non-unitary effective evolution. {Under weak-coupling and Markov assumptions this may take a Lindblad form. With the additional relaxational projector closure discussed in Appendix~\ref{app:lindblad}, we model the slow sector by}
\begin{equation}
\partial_\tau P = -[P,[P,{\cal H}_{\rm eff}[P]]] + \eta.
\end{equation}
The stochastic noise $\eta$ is thus interpreted as the effective imprint of decoherence induced by unresolved GFT degrees of freedom acting as an environment for the emergent geometric sector.

\section{Conclusions}

\begingroup
We have formulated an effective framework connecting correlation geometry, frame transport, and a constrained stochastic projector flow. Smooth fixed-rank projectors define a Wilczek--Zee connection, and refined cyclic projector overlaps recover its traced parallel transport.

Assuming a local continuum regime and a suitable massive fluctuation spectrum, the one-loop derivative expansion contains Yang--Mills-type and Einstein--Hilbert-type terms. This provides an induced-action mechanism within the effective description, rather than a derivation of continuum spacetime or its signature from the measure alone.

The frame action also supplies a Grassmannian symplectic structure. Its identification with the gravitational Ashtekar--Barbero phase space requires additional variables, a nondegenerate triad, symplectic matching, and the gravitational constraint algebra. A formal pullback to connection variables does not establish these properties.

The geometric, disordered, and topological regimes discussed here are candidate organizations of the correlation sector. Demonstrating their existence, confinement diagnostics, or phase transitions requires a specified action and a controlled large-system limit. Deriving the stochastic coefficients and their relation to coarse-graining from an explicit microscopic model is likewise an open task.

These distinctions delimit the present results while providing a concrete programme for investigating emergent spacetime and interactions through correlation geometry.
\endgroup

\begin{acknowledgments}
\noindent 
{We are indebted to Claudio Paganini for insightful comments.} 
\end{acknowledgments}

\appendix

\section{One-loop induced gauge action}
\label{app:oneloop}
\noindent 
{We illustrate the induced gauge term in an auxiliary local scalar model with a background connection. Identifying its massive modes and measure with those of a constrained correlation model is a separate matching assumption.}

\subsection{Linear sigma-model formulation}
\noindent 
Consider an $N\times k$ complex matrix field $\Phi$ with action
\begin{eqnarray}
S[\Phi,A]
=
\int d^dx \,
\big[ && \!\!\!\!
\Tr\big((D_\mu\Phi)^\dagger D^\mu\Phi\big)
+
m^2 \Tr(\Phi^\dagger\Phi) \nonumber\\
+ && \!\!\!\!
\lambda \Tr\big((\Phi^\dagger\Phi-v^2\mathbb{I})^2\big)
\big],
\end{eqnarray}
with
\begin{equation}
{D_\mu\Phi = \partial_\mu\Phi + i\Phi A_\mu.}
\end{equation}

In the large-$\lambda$ limit, $\Phi^\dagger\Phi\simeq v^2\mathbb{I}$ and $\Phi=v\Psi$.

\subsection{Functional determinant}
\noindent 
{At Gaussian order, $n_s$ massive complex scalar multiplets in representation $R$ contribute}
\begin{equation}
S_{\rm eff}[A]
=
n_s\, \Tr\log(-D^2+M^2),
\end{equation}
{Here $n_s$ counts complex multiplets, and $\tr_R(T^aT^b)=T(R)\delta^{ab}$. The determinant is a quadratic-fluctuation result, not the exact integral of the interacting sigma model.}

Using the proper-time representation,
\begin{equation}
\Tr\log(-D^2+M^2)
=
{-\int_{\Lambda_{\rm UV}^{-2}}^\infty \frac{ds}{s} e^{-sM^2}} \Tr(e^{sD^2}),
\end{equation}
and the heat-kernel expansion,
\begin{equation}
\Tr(e^{sD^2})
=
\int d^dx \frac{1}{(4\pi s)^{d/2}}
\tr\left[
\mathbb{I}
{-
\frac{s^2}{12}F_{\mu\nu}F^{\mu\nu}}
+ \cdots
\right],
\end{equation}
{where the minus sign follows from the anti-Hermitian bundle curvature $\Omega_{\mu\nu}=[D_\mu,D_\nu]$ and $\tr\Omega^2=-\tr F^2$ \cite{Vassilevich:2003}. Combining this with the proper-time minus sign, we obtain}
\begin{equation}
S_{\rm eff}[A]
\supset
\frac{1}{4g_{\rm eff}^2}
\int d^dx\, F_{\mu\nu}^a F^{a\mu\nu}.
\end{equation}

In $d=4$,
\begin{equation}
{\left.\frac{1}{g_{\rm eff}^2}\right|_{\rm induced}
=\frac{n_sT(R)}{48\pi^2}
\log\frac{\Lambda_{\rm UV}^2}{M^2}+O(1).}
\end{equation}

\section{Induced Einstein--Hilbert term}
\label{app:EH}
\noindent 

{We compute the curvature term conditional on a smooth Riemannian metric and a stable Laplace-type fluctuation operator. This does not reconstruct a metric from projectors or derive Lorentzian signature.}

In the geometric phase of the stochastic flow, slowly varying projector configurations admit an effective continuum description in terms of emergent metric and connection variables. The corresponding infrared transport structure defines an effective background geometry, analogous to the emergent continuum metrics appearing in induced-gravity {approaches~\cite{Sakharov:1968}}.

Within this effective geometric regime, fluctuations of the correlation frame are governed by a Laplace-type operator
\begin{equation}
\Delta
=
-g^{\mu\nu}D_\mu D_\nu
+
\mathcal E ,
\end{equation}
{where $g_{\mu\nu}$ is the assumed infrared Riemannian metric, the second derivative includes its Levi--Civita connection, $\mathcal E$ is an endomorphism, and}
\begin{equation}
D_\mu=\nabla_\mu-iA_\mu
\end{equation}
is the corresponding covariant derivative acting on the fluctuation modes.

The one-loop effective action generated by integrating out the fast correlation-sector fluctuations is then
\begin{equation}
S_{\rm eff}
=
-
n_s
{\int_{\Lambda_{\rm UV}^{-2}}^\infty
\frac{ds}{s}}
\,e^{-sM^2}
\,
{\rm Tr}
\left(
e^{-s\Delta}
\right).
\end{equation}

The heat-kernel expansion reads
\begin{equation}
\Tr(e^{-s\Delta})
=
\int d^4x \sqrt{g}\,
\frac{1}{(4\pi s)^2}
\tr\left[a_0+s\,a_1+s^2a_2+\cdots\right],
\end{equation}
with
\begin{equation}
a_0=\mathbb I,
\qquad
a_1=\frac16 R\,\mathbb I - \mathcal E.
\end{equation}
Substituting into the proper-time integral gives
\begin{equation}
S_{\rm eff}
\supset
-\frac{n_s}{(4\pi)^2}\int d^4x \sqrt{g}
\left[
c_0 \Lambda_{\rm UV}^4
+
c_1 \Lambda_{\rm UV}^2 R
+\cdots
\right],
\end{equation}
\begingroup
For example, for $n_s$ complex scalar multiplets of dimension $r$ with $\mathcal E=\xi R\mathbb I$, the coefficient of $\int\sqrt g\,R$ is
\begin{equation}
c_R=-\frac{n_s r(1/6-\xi)}{(4\pi)^2}
\int_{\Lambda_{\rm UV}^{-2}}^\infty\!ds\,s^{-2}e^{-sM^2}.
\end{equation}
This coefficient has a quadratic cutoff dependence and mass-dependent terms; finite counterterms and the renormalization prescription affect its value. For the Euclidean convention $S_E\supset-(16\pi G)^{-1}\int\sqrt g\,R$, one identifies $c_R=-(16\pi G)^{-1}$ only after these choices. Neither a universal positive Newton constant nor the value $n_sM^2/(96\pi^2)$ follows without specifying the spectrum, curvature couplings, and matching conditions.
\endgroup

\section{Stochastic projector dynamics from Lindblad evolution}
\label{app:lindblad}
\noindent 
{This Appendix motivates the projector-flow ansatz using open-system dynamics and states the extra assumptions required for a relaxational closure. A general Lindblad equation does not uniquely determine this projector process.} {For a model-dependent realization of the connection between stochastic geometric evolution and a Lindblad description, see Ref.~\cite{Lulli:2023srf}, especially Eq.~(23) and Appendix~F.} The key point is that the slow geometric sector is not closed: it interacts with unresolved microscopic GFT modes, graph-rearrangement channels, and other fast degrees of freedom which act as an effective environment. After tracing out these modes, the reduced dynamics of the coarse-grained sector becomes non-unitary and, under standard assumptions, admits a Lindblad description.

\subsection{Reduced dynamics of the correlation sector}
\noindent 
We decompose the total Hilbert space as
\begin{equation}
{\cal H}_{\rm tot}
=
{\cal H}_{\rm corr}\otimes {\cal H}_{\rm env},
\end{equation}
where ${\cal H}_{\rm corr}$ denotes the slow correlation sector and ${\cal H}_{\rm env}$ the unresolved microscopic environment. {Writing $s$ for open-system evolution time, distinct from the auxiliary parameter $\tau$, let the total density matrix obey}
\begin{equation}
{\partial_s \rho_{\rm tot}}
=
-i[H_{\rm tot},\rho_{\rm tot}],
\qquad
H_{\rm tot}=H_{\rm corr}+H_{\rm env}+H_{\rm int}.
\end{equation}
The reduced density matrix
\begin{equation}
\rho_{\rm corr}=\Tr_{\rm env}\rho_{\rm tot}
\end{equation}
then evolves non-unitarily. In the weak-coupling, Markovian, and coarse-grained limit one obtains a Lindblad master equation,
\begingroup
\begin{align}
\partial_s\rho_{\rm corr}
&=-i[H_{\rm eff},\rho_{\rm corr}]\nonumber\\
&\quad+\sum_\alpha\left(
L_\alpha\rho_{\rm corr}L_\alpha^\dagger
-\tfrac12\{L_\alpha^\dagger L_\alpha,\rho_{\rm corr}\}\right).
\label{eq:LindbladAppendix}
\end{align}
\endgroup

where $H_{\rm eff}$ is the environment-dressed effective Hamiltonian and the $L_\alpha$ encode dissipative couplings to the unresolved microscopic sector.

\subsection{Projector observables}
\noindent 
Let $P(x,\tau)$ denote the coarse-grained local correlation projector. At the operator level, its expectation value is
\begin{equation}
{p_P(x,s)=\Tr\!\big(\rho_{\rm corr}(s)\,P(x)\big).}
\end{equation}
Using Eq.~\eqref{eq:LindbladAppendix}, one finds
\begin{align}
{\partial_s p_P}
&=
-i\,\Tr\!\big(\rho_{\rm corr}[P,H_{\rm eff}]\big)
\nonumber\\
&\quad
+
\sum_\alpha
\Tr\!\left(
\rho_{\rm corr}
\Big(
L_\alpha^\dagger P L_\alpha
-\frac12\{L_\alpha^\dagger L_\alpha,P\}
\Big)
\right).
\label{eq:meanP}
\end{align}
{The first term gives coherent evolution and the second the dissipative contribution. The quantity $p_P$ is a scalar expectation value, not a projector; its equation is not a closed equation for $P$.}

To connect with the geometric flow used in the main text, we now restrict the dynamics to the slow manifold of projector configurations. A tangent variation of a projector satisfies
\begin{equation}
\delta P = [P,\Omega],
\qquad
\Omega^\dagger=-\Omega,
\end{equation}
so the natural projector onto the tangent space of the Grassmannian is
\begin{equation}
\Pi_P(X)=[P,[P,X]].
\end{equation}
{To obtain the drift used in the main text we additionally assume an overdamped closure with a Hilbert--Schmidt gradient of a chosen effective free energy and no retained coherent drift. Under this assumption,}
\begin{equation}
\partial_\tau P\Big|_{\rm drift}
=
-\Pi_P\!\left(\frac{\delta {\cal F}[P]}{\delta P}\right),
\label{eq:driftappendix}
\end{equation}
{where ${\cal F}[P]$ is the effective free-energy functional of the closure; its derivation from specified $L_\alpha$ is not given here.} Equivalently, one may write
\begin{equation}
\partial_\tau P\Big|_{\rm drift}
=
-[P,[P,{\cal H}_{\rm eff}[P]]],
\end{equation}
{with ${\cal H}_{\rm eff}[P]=\delta{\cal F}/\delta P$ the gradient generator, distinct from the Hamiltonian $H_{\rm eff}$ in the master equation.}

\subsection{From Lindblad evolution to stochastic trajectories}
\noindent 
The Lindblad equation for $\rho_{\rm corr}$ admits an unraveling in terms of stochastic pure-state trajectories. {An unraveling is not unique, and its rank-one state projectors need not equal the coarse-grained rank-$k$ spectral projectors. Assuming a separate closure on the latter, we model their evolution by the Stratonovich process}
\begin{equation}
dP_\tau
=
-\Pi_{P_\tau}\!\left(\frac{\delta {\cal F}[P]}{\delta P}\right)d\tau
+
{\sum_\alpha \Pi_{P_\tau}\!\big({\cal B}_\alpha[P_\tau]\big)\circ dW_\tau^\alpha,}
\label{eq:stochasticPgeneral}
\end{equation}
{where the $dW_\tau^\alpha$ are independent real Wiener increments and the Hermitian noise fields ${\cal B}_\alpha[P]$ are closure data to be matched to a microscopic model. Tangent vector fields preserve the constraints in Stratonovich form. An equivalent It\^o equation requires the additional drift $\tfrac12\sum_\alpha (Db_\alpha)_P[b_\alpha]$, with $b_\alpha=\Pi_P({\cal B}_\alpha)$.}

For an isotropic coarse-grained environment, the noise may be approximated by an effective Gaussian matrix-valued process, yielding
\begin{equation}
dP_\tau
=
-[P_\tau,[P_\tau,{\cal H}_{\rm eff}[P_\tau]]]\,d\tau
+
\sqrt{2D}\,\Pi_{P_\tau}\circ dW_\tau,
\label{eq:stochasticPfinal}
\end{equation}
where $D$ is an effective diffusion coefficient and $\circ$ denotes Stratonovich multiplication. In differential form this is the stochastic projector equation used in the main text,
\begin{equation}
\partial_\tau P
=
-[P,[P,{\cal H}_{\rm eff}[P]]]
+\eta,
\end{equation}
with $\eta$ the stochastic force induced by unresolved microscopic GFT modes.

{Equation~\eqref{eq:stochasticPfinal} is therefore an open-system-motivated effective ansatz. Its noise amplitude, free energy, and relation between $s$ and $\tau$ require an explicit microscopic calculation.} The deterministic double-commutator term describes relaxation along the slow manifold of projector configurations, while the noise term encodes fluctuations generated by the environment. In this way, the stochastic gauge--geometry flow of the main text is naturally interpreted as the open-system dynamics of emergent correlation sectors.\\

In the GFT setting, the environment ${\cal H}_{\rm env}$ may be understood as comprising graph-refinement modes, local adjacency-rearrangement channels, and microscopic topology-changing excitations which are not explicitly resolved in the effective continuum description. Their net effect on the slow entanglement-defined geometric sector is captured by the Lindblad operators $L_\alpha$.

\section{Reduction of the frame symplectic structure}
\label{app:symplectic}
\noindent 
The frame field $\Psi(x)$ defines a point in the Stiefel manifold of orthonormal $k$-frames,
\begin{equation}
\Psi^\dagger \Psi = I_k ,
\end{equation}
while the corresponding projector
\begin{equation}
P=\Psi\Psi^\dagger
\end{equation}
defines a point in the Grassmannian of $k$-dimensional subspaces.

The map
\begin{equation}
\Psi \mapsto P
\end{equation}
defines a principal $U(k)$ bundle, whose fibers correspond to local frame rotations
\begin{equation}
\Psi(x)\rightarrow \Psi(x)V(x).
\end{equation}

The first-order frame action
\begin{equation}
S_{\rm frame}
=
\int dt \int_\Sigma d^3x
\left(
{i\kappa\,{\rm Tr}(\Psi^\dagger\partial_t\Psi)}
-
H_{\rm frame}
\right)
\end{equation}
defines the symplectic potential
\begin{equation}
\Theta_{\rm frame}
=
i\kappa
\int_\Sigma d^3x\,
{\rm Tr}(\Psi^\dagger \delta\Psi).
\end{equation}

Variations naturally decompose into vertical and horizontal components,
\begin{equation}
\delta\Psi
=
\delta\Psi_\perp
+
\Psi\alpha ,
\end{equation}
where
\begin{equation}
\alpha=\Psi^\dagger\delta\Psi
\end{equation}
generates infinitesimal $U(k)$ frame rotations.

The horizontal component satisfies
\begin{equation}
\Psi^\dagger \delta\Psi_\perp =0 ,
\end{equation}
and therefore corresponds to genuine deformations of the correlation subspace rather than changes of basis.

{The two-form $\delta\Theta_{\rm frame}$ descends under this quotient, while $\Theta_{\rm frame}$ itself is a local potential and is not horizontal. In projector variables, the descended form is $\Omega_{\rm Gr}=i\kappa\int_\Sigma d^3x\,\Tr(P\,\delta P\wedge\delta P)$. The first-order action is an additional ansatz; global consistency of its Berry phase must be imposed separately.}\\

In the infrared geometric regime, we assume that the slowly varying projector field $P(x)$ admits a local continuum description in terms of an emergent connection
\begin{equation}
{A_a^i(x)
=
2i\,{\rm Tr}
\left(
\tau^i \Psi^\dagger \partial_a\Psi
\right).}
\end{equation}

{The composite-connection map need not be injective, surjective, or a system of independent coordinates on the reduced phase space. The following pullback is conditional on choosing an admissible local family; it does not establish such a parametrization for arbitrary connections.}

Restricting the frame field to a family
\begin{equation}
\Psi=\Psi[A]
\end{equation}
parameterized by the emergent connection defines a pullback of the frame symplectic potential onto the reduced configuration space,
\begin{equation}
\Theta_{\rm red}[A]
=
i\kappa
\int_\Sigma d^3x\,
{\rm Tr}
\left(
\Psi^\dagger[A]\,
\delta\Psi[A]
\right).
\end{equation}

\section{Explicit pullback of the frame symplectic potential}
\label{app:pullback}
\noindent 
Using
\begin{equation}
\delta\Psi(y;A)
=
\int_\Sigma d^3x\,
\frac{\delta\Psi(y;A)}
{\delta A_a^i(x)}
\,
\delta A_a^i(x),
\end{equation}
one obtains
\begin{equation}
\Theta_{\rm red}
=
\int_\Sigma d^3x\,
E_i^a(x;A)\,
\delta A_a^i(x),
\end{equation}
where
\begin{equation}
E_i^a(x;A)
=
\frac{i\kappa}{2}
\int_\Sigma d^3y\,
{\rm Tr}
\left[
\Psi^\dagger
\frac{\delta\Psi}{\delta A_a^i(x)}
-
\frac{\delta\Psi^\dagger}{\delta A_a^i(x)}
\Psi
\right]
\end{equation}
{is the coefficient of $\delta A$ in this chosen pullback. This definition alone is not a moment-map construction and does not establish an independent gravitational momentum.}\\

\begingroup
Taking an exterior derivative gives the formal identity
\begin{equation}
\delta\Theta_{\rm red}
=\int_\Sigma d^3x\,\delta E_i^a[A]\wedge\delta A_a^i.
\end{equation}
Because $E[A]$ depends on $A$, this is not the canonical two-form on a space of independent connections and triads. Its nondegeneracy and the admissible variations must be determined before defining Poisson brackets.

No local triad formula is asserted here. In particular, $P$ acts on the $N$-dimensional ambient space, whereas $A$ and $\tau_i$ act on the $k$-dimensional frame fiber. Expressions such as $[A_a,P]$ or $\Tr(\tau_i D_bP D_cP)$ are not defined without an additional ambient embedding and cannot establish the missing canonical identification. A dimensionally consistent triad construction, the full symplectic map, and the gravitational constraints are left for further work.
\endgroup

%\clearpage
\bibliographystyle{apsrev4-1}
\bibliography{references}

\end{document}